\documentclass[final,12pt]{elsarticle}
\usepackage[margin=2.3cm]{geometry}
\usepackage{graphics}
\usepackage{graphicx}
\usepackage{epsfig}

\usepackage{amssymb}
\usepackage{amsthm}

\usepackage{changes}
\usepackage{amsmath}
\usepackage{hyperref}
\usepackage{slashbox}
\usepackage{url}
\usepackage{setspace}
\usepackage{subfiles}
\usepackage{natbib}
\usepackage{csquotes}
\usepackage[caption=false,font=footnotesize]{subfig}
\usepackage{caption}
\usepackage{booktabs}
\usepackage{tabularx}
\usepackage{multirow}
\usepackage{algorithm}
\usepackage{algorithmic}
\journal{Reliability Engineering and System Safety}
\usepackage{hyperref}
\pdfoutput=1
\begin{document}

\begin{frontmatter}

\title{Robustness Assessment of Hetero-Functional Graph Theory Based Model of Interdependent Urban Utility Networks}

\author{Sai Munikoti}
\author{Kexing Lai}
\author{Balasubramaniam Natarajan}
\address{Department of Electrical and Computer Engineering,  Kansas State University, Manhattan, KS 66506 USA. saimunikoti@ksu.edu, klai@ksu.edu, bala@ksu.edu}

\begin{abstract}
The increasing urban population imposes a substantial and growing burden on the supporting infrastructure, such as electricity, water, heating, natural gas, road transportation, etc. This paper presents a Hetero-functional graph theory (HFGT) based modeling framework for these integrated infrastructures followed by an analysis of network robustness. The supporting infrastructures along with the infrastructure repair facilities are considered. In contrast to conventional graph representations, a weighted HFGT model is used to capture the system processes and mutual dependencies among resources. To assess robustness of the inter-dependent networks, impacts of complete/partial and random/targeted attacks are quantified. Specifically, various contingency scenarios are simulated and the vulnerability of the network is evaluated. Additionally, several robustness metrics are proposed to provide a comprehensive evaluation of system robustness. 
The proposed weighted HFGT modeling and robustness assessment approach is tested using a synthetic interdependent network, comprising of an electrical power system, a water network, a district heating network, a natural gas system and a road transportation network. Results demonstrate that system robustness can be enhanced via securing system information and mitigating attack strength. 
\end{abstract}


\end{frontmatter}

\section{Introduction}
\subsection{Motivation and Background}

The world population has increased by 113.3\% \cite{WorldBank} in the past 50 years. More importantly, with massive urbanization, the percentage of global population in urban areas increased from 34\% to 52\% in the past five decades \cite{WorldBankUrPo}. The swelling urban population imposes increasingly intense demands over a variety of infrastructures. A large-scale urban area is comprised of multiple utility networks that deliver a variety of services to the community, including electricity \cite{Talaat2020}, water \cite{Oikonomou2019}, heating \cite{Zajacs2019}, natural gas \cite{ZhaoUC}, road transportation \cite{Mahmoudi2019}, etc. These networks are inherently integrated and their interdependencies are becoming increasingly tight due to the increased reliance on cyber infrastructure to enable smart and efficient operation.

The concepts of resilience and robustness have attracted tremendous attention in recent years in response to the frequent and widespread natural disasters and man-made malicious attacks across the globe. These extreme events impose significant threats to our urban infrastructures \cite{liu2020review}, \cite{Najafi2018}, \cite{das2020measuring}. Several approaches have been proposed to model and evaluate the robustness of interdependent urban infrastructures such as hierarchical graph representation \cite{ferrario2016evaluation}, dynamic network flow model \cite{goldbeck2019resilience}, agent based modeling \cite{dubaniowski2020framework}, among others. This paper, for the first time, proposes adopting a ``weighted hetero-functional graph theory'' concept \cite{schoonenberg2017modeling} to model and conduct robustness analysis of interdependent urban infrastructure networks, comprising of electricity, water, heating, natural gas and road transportation networks (along with infrastructure repair services). 

\subsection{Literature Review}

The inter-dependencies of multiple infrastructures and services call for coordination of the constituent systems. For instance, the increasing use of natural gas for electricity production embodies inter-dependencies between natural gas and electric power infrastructures. Accordingly, many research works have been conducted to investigate the coordination and co-optimization of the two systems. For instance, Zhao et al. carried out a series of research works on coordinated planning \cite{ZhaoPlan} and operation \cite{ZhaoUC} of integrated gas-electricity systems. It was found that strategic coordination would aid in addressing several practical challenges in natural gas systems, including natural gas leakage \cite{ZhaoStorage}, rapid changes and significant uncertainties in gas demand \cite{ZhaoPlan}, and natural gas-supply bottlenecks \cite{ZhaoUC}, etc. Additionally, integrated gas-electricity planning and operation is helpful in tackling the challenges frequently witnessed in electric power systems, such as (1) the substantial gap between peaks and valleys of demand profiles \cite{Gholizadeh2019}, (2) the inherently stochastic characteristics of renewable production \cite{Ansari2020}, (3) the poor performance of electricity demand prediction \cite{Obringer2020}, etc. Furthermore, the unwillingness of both natural gas and electricity system operators to share proprietary information is taken into account in numerous studies. Accordingly, optimization frameworks for joint gas-electricity operation with limited information exchanges are proposed to mitigate potential barriers of coordinating the two systems \cite{ZhaoCoo}.  

Apart from gas-driven power plants that embody the integration between natural gas and power networks, gas based combined heat and power (CHP) technology exploits gas engines to generate electricity and useful heat simultaneously \cite{Li2018}. In the United States, the efficiency of CHP applications is expected to reach 65\% - 75\%, which exhibits a significant improvement over the national average of 50\% efficiency if heating and electricity delivery services are provided separately \cite{DOECHP}. As a result, CHP technology allows for the tight integration of gas, electricity and heating networks. To explore approaches to promote such integration, \cite{Li2018} proposed stochastic optimal operation of a low-carbon micro integrated electric power, natural gas and heating delivery systems, centered around CHP facilities. \cite{Clegg2016} proposed a novel framework for modeling flexibility of integrated gas, electricity and heating network with gas-driven CHP technology. A unified steady-state energy flow analysis considering electrical, natural gas and heating networks with gas-driven CHP facilities as the coupling points was presented in \cite{Shabanpour-Haghighi2016}. Excluding the electrical sector, heating and cooling sector alone accounts for a large portion of the energy consumption portfolio (with 50\% of European Union's annual energy consumption in 2019) \cite{Dominkovic2020}. Further, in the European Union, the largest energy source for heating and cooling is natural gas with 46\% share \cite{Dominkovic2020}. In addition to the CHP technology mentioned earlier, natural gas powered heating/cooling plants serve as an alternative source for district heating networks (DHN) \cite{Dorfner2014}. As a consequence, co-optimization of integrated gas-electricity-heating networks incorporating power plants, heating plants and CHP facilities, which are all driven by natural gas, has been studied in prior literature, such as \cite{Catrini2020}, \cite{tian2019large}.   

Electricity infrastructures and water distribution system (WDS) facilities are also tightly integrated. In the United States, WDS facilities, including water plants and water pipelines, are energy intensive infrastructures that account for 3\% - 4\% of the total electricity consumption \cite{Oikonomou2019}. Multiple prior publications have discussed the inter-dependency and proposed the schemes to coordinate these two infrastructures \cite{Lee2020}. For instance, \cite{Oikonomou2019} proposed a network-constrained unit commitment model for electric power systems, considering flexibility of electricity usage for WDS. \cite{Lee2020} applied a Bayesian network to estimate the service disruptions of integrated power and water supply system after an earthquake. This analysis provides insights into the resilience of electricity-water network infrastructures. \cite{Najafi2020} suggested the use of private-owned microgrids and water pumps for enhancing accessibility to water and power after natural disasters. 
Aside from power lines and pipelines that interconnect electricity, heating, gas and water facilities, road transportation systems act as an indispensable sector that delivers a variety of materials, services and merchandise to those infrastructures. Incorporating the geographical information and traffic conditions into energy system operation has been discussed in \cite{CarSharing2020}.

In this paper, for the first time, interdependent urban infrastructure networks (IUNs) including electricity, natural gas, heating, water and road transportation for repair service are modeled under a common framework. A weighted hetero-functional graph theory (HFGT) \cite{schoonenberg2017modeling} is used as the modeling framework within which robustness assessment is conducted. HFGT can be viewed as an intellectual fusion of model-based systems engineering and network science, and has been applied in transportation networks \cite{Wardt2017}, production systems \cite{Schoonenberg2017}, power grids \cite{Farid2015}, among others. Along with capturing physical dependency, it also models a functional relationship. Specifically, seven models can be used to study network component dependency, including (1) system concept, (2) hetero-functional adjacency matrix, (3) controller agency matrix, (4) controller adjacency matrix, (5) service as operand behavior, (6) service feasibility matrix, and (7) system adjacency matrix. In this paper, we focus on the system concept model. Compared to conventional multi-layer graphs \cite{Park2019}, the system concept model provides structural descriptions of a system that indicates the mapping of system resources to system functionalities (processes). Similarly, the process relation model has been proposed which relates the dependency among various processes. For example, in a power system network, the graphical nodes and edges in the process relation model represent the processes and dependencies between processes, instead of buses and power lines \cite{Assadian2011}. The process relation model can therefore reveal in-depth information about system configurations and capabilities. Further, various attack simulations can be conducted on the process relation graphical model to evaluate robustness. It is worth to note that the ``attack'' word used throughout this paper, represents contingencies (complete or partial loss of nodes) in the network that can be caused by both malicious attacks and natural disasters.

\subsection{Contributions}

The major contributions of this paper are listed below. 
\begin{itemize}
	\item This paper, for the first time, introduces a weighted Hetero-functional graph theory based approach to model urban interdependent infrastructure networks. Unlike \cite{schoonenberg2017modeling}, which merely presents a visual representation, this paper conducts an in-depth analysis that reveals system robustness against contingencies from multiple dimensions.   
	
	\item System network configuration information is further incorporated into Hetero-functional graph development. Specifically, the dependencies and interconnections among various processes due to network constraints are modeled in detail, enhancing the practical applicability of this work.
	
	\item A comprehensive evaluation of the robustness of weighted IUN is presented for the first time. Four types of attacks with various strengths and knowledge levels are modeled and simulated. These attacks include (1) complete random attacks; (2) complete targeted attacks; (3) partial random attacks; and (4) partial targeted attacks. In other words, instead of simply presuming a process will be entirely disabled after an attack, we cover a variety of real-world attack scenarios that might only lead to partial degradation of the associated process until it becomes completely dysfunctional. Therefore, the proposed assessment strategy would provide realistic and inclusive suggestions for decision makers to optimize the dependency relations among infrastructures for robustness enhancement.    
\end{itemize}

The remainder of this paper is organized as follows. Section 2 presents an overview of conventional and HFGT models with proposed weighted HFGT framework. Section 3 discusses robustness evaluation and various attack strategies. In Section 4, numerical results are presented along with the comparison across attack strategies. Section 5 presents the conclusions.   

\section{Modeling Integrated Urban Networks}

This section presents existing paradigms for modeling IUN, followed by an overview of the proposed framework. Specifically, the system knowledge base of HFGT that maps the allocation of system resources to system processes will be depicted and elaborated here.

\subsection{Conventional Graph Theory framework}

\begin{figure}[!ht]
	\centering
	\includegraphics[width=0.98\textwidth]{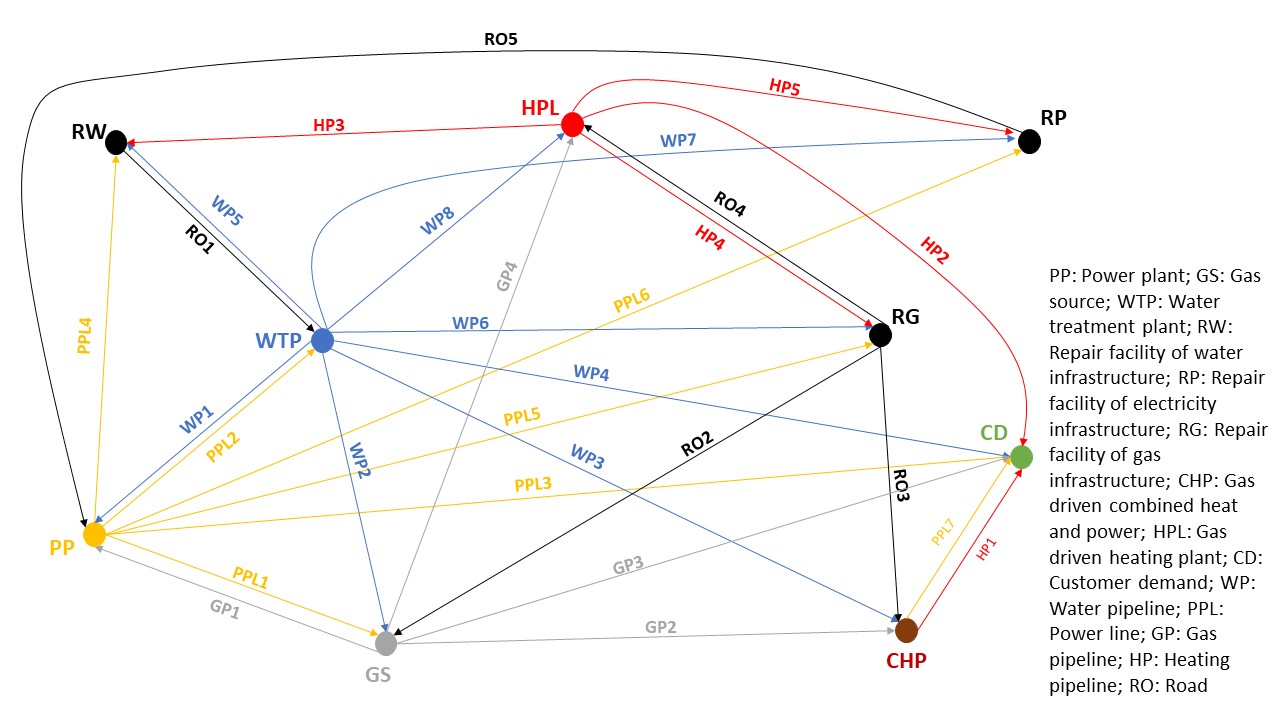}
	\caption{Conventional Graph Representation of a Conceptual Interdependent Urban Infrastructure Network (IUN)}
	\label{Fig:ConGraNet}
\end{figure}

A conceptual diagram for an IUN is depicted in Fig. \ref{Fig:ConGraNet}, which is used as an example to illustrate the interactions among infrastructures. In this directed graph representation, the nodes denote various infrastructures and facilities, while directed edges represent the energy/water/service flow among the nodes. As seen, networks of electricity, natural gas, heating, and water are intertwined to support each other. For instance, the operative gas-driven power plant (PP1) depends on the gas supply from the gas source (GS) and water supply from the water treatment plant (WTP). Besides, the effective maintenance and repair service for PP1 relies on the power infrastructure repair facility (RP). On the other hand, PP1 provides electricity to other infrastructures and facilities, such as GS, WTP, among others. Additionally, customer demands (CD) for energy and water supplies are supported by the corresponding facilities.   

\subsection{Hetero-functional Graph Theory (HFGT) framework}

Although the conventional graph representation can reveal the topological information and fundamental relationship among system components, for complex systems, it does not provide in-depth knowledge about system processes and mutual dependencies among resources. In other words, questions such as \emph{``What are the dependencies of a particular facility on the others?"}, \emph{``What processes are performed within nodes and edges?"}, \emph{``How important are nodes and edges for system performance?"} cannot be properly addressed by analyzing conventional graph representations.   
Therefore, Farid et al. \cite{schoonenberg2017modeling} introduced an HFGT framework to rigorously model IUN and tackle all the challenges unaddressed within the conventional graph theory paradigm. The skeleton of HFGT consists of seven building blocks starting with the system concept model that maps system processes to system resources. System process is defined as an activity that transforms a predefined set of inputs into a predefined set of outputs, e.g. generating electricity from a PP, producing treated water from a WTP, etc. System resource is a physical entity which executes the processes, such as PP, WTP, GS, etc. The system concept model provides a rigorous platform to conduct this transformation. Specifically, the system knowledge matrix ($J_S$) in the system concept model is a binary matrix, which maps the processes ($P$) with the corresponding resources ($R$) as follows: 

\begin{equation}
\label{SCMap}
P =  J_{S}\odot R 
\end{equation}
where $\odot$ is matrix boolean multiplication \cite{schoonenberg2019hetero}. The element in the $J_S$ that projects the element in $R$ to the corresponding element in $P$ is equal to unity, when an action that converts the resource to the process exists. For instance, a power plant, as a system resource, can be mapped to generating electricity, as a system process, via the knowledge matrix ($J_S$). Accordingly, the particular element in $J_S$ representing the electricity generation action being executed by the power plant is set to unity. 


\subsection{Weighted hetero functional graph theory (WHFGT) framework}
\begin{figure}[!ht]
	\centering
	\includegraphics[scale=0.5]{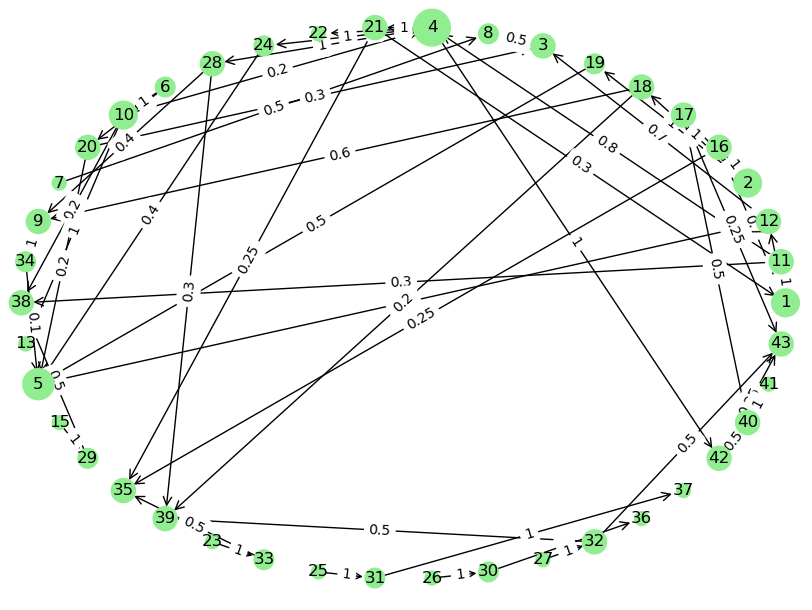}
	\caption{Process relation graph corresponding to the network shown in Fig. 1}
	\label{Fig:Prg}
\end{figure}

Using the knowledge matrix, system resources and system processes can be mapped. To more realistically capture process interactions, we propose a process relation matrix ($P_R$), which is developed on top of the conventional dependency matrix of HFGTs. For the conventional dependency matrix, the relationships between the processes are expressed in binary terms, i.e., it only captures whether the dependency is present or not. However, it is inadequate to quantify the degree of dependency between the two processes. Therefore, to address this shortcoming, we propose a weighted HFGT (WHFGT) framework and the corresponding relation matrix ($P_R$), which quantifies the degree of dependency using real valued weight. For a particular node in the WHFGT, the weights of its associated edges in matrix $P_R$ denote the degree of dependencies on other processes. For instance, in Fig. \ref{Fig:ConGraNet}, electricity demand of the community customers can be supplied by two sources, i.e., the gas-driven power plant and CHP, and the dependencies on both of them can be reflected by assigning weights. The process relation matrix ($P_R$) for the conceptual diagram shown in Fig. \ref{Fig:ConGraNet} is shown in \cite{ProcessMatrix1}, which is a $43\times43$ matrix as $43$ processes are incorporated. Correspondingly, Fig. \ref{Fig:Prg} depicts the process relation graph for a synthetic network shown in Fig. \ref{Fig:ConGraNet}. 
The direction of edges are from the source to target nodes, where target nodes (processes) are dependent on the source nodes (processes) with dependency proportional to the edge weights. In the following sections, extensive analysis will be conducted on the process relation graph to evaluate robustness of IUN. The WHFGT allows one to study the interdependency in a more informative way in a sense that the dependency weights could play an important role in various graph robustness analysis (percolation, cascading failures, etc.). 

\section{Contingency Analysis in WHFGT Framework}
The tight interconnections among various infrastructures may exacerbate the impact of potential cascading failures, as contingencies in one part of the system might propagate to other networks. Therefore, assessment of robustness for interdependent urban networks is critical for vulnerability analysis and robustness enhancement. 

To simulate contingencies on IUNs, the nodes are defined as the targets for attacks. In other words, disabling certain processes is equivalent to removing the associated nodes in the WHFGT graphs. Besides, based on graph percolation procedure \cite{callaway2000network}, nodes are removed in stages and the robustness metrics of the residual graph are computed at each stage to study the transition properties of the graph. Along with the removal of a selected node, the child nodes of the selected node are also removed under specific conditions. Here, child nodes are defined as nodes that can be reached (directed path exist) from the selected node and are computed using depth first search algorithm \cite{cormen2009introduction}. We would like to reiterate that, for the sake of simplicity, all the contingencies in this paper are referred to as \enquote{attacks}.

According to the available system information and attacker capabilities/strength, the attacks can be further divided into four categories, as tabulated in Table \ref{Table:Typesofattack}. Random/targeted attacks are categorized based on the knowledge level of attackers about the system. Specifically, in a random attack, nodes are removed randomly at each stage of the percolation whereas, in a targeted attack, nodes are removed based on certain importance scores of nodes (assuming the system information is available to the attacker). On the other hand, complete/partial attacks are categorized based on attackers capability. Specifically, in a complete attack, the attacker disables a process completely, while in a partial attack, the attack can only degrade the process to a certain degree. Detailed description of these attack categories will be provided later.

\subsection{Graph Centrality}
As mentioned earlier, targeted attacks are launched based on the importance scores assigned to nodes, assuming the attacker has acquired the entire system information. Here, the concept of centrality is applied to quantify importance of nodes in a network. Four types of centrality are widely applied in prior literature, including degree centrality, eigenvector centrality, pagerank centrality and betweenness centrality \cite{gao2013modified}. The degree centrality indicates the number of connected edges for each node. The intuitive interpretation of degree is that nodes with more connections imply involvement in more functionalities and transformations. Moreover, for directed graphs, degree centrality can be separated into in-degree and out-degree, which represent the number of incoming edges and outgoing edges, respectively. Thus, weighted out-degree denotes the total sum of outgoing edge weights. Eigenvector centrality provides an alternative perspective of the node's importance. Here, the focus is on quantifying the influence of connected nodes, instead of using the number of connected nodes. A potential problem faced by eigenvector centrality is that if one node is among a large number of nodes connected to another node with substantially high centrality, the eigenvector centrality index would overestimate the influence of that particular node. PageRank centrality is proven to be able to resolve this issue and is one of the most well-known and frequently used measure in search engines \cite{Tu2018}. Finally, the betweenness centrality measures the extent to which a node lies on the shortest paths between other nodes. Nodes with high betweenness are expected to have considerable influence within a network by virtue of their control over information passing between others. In IUN, we select betweenness and weighted out-degree centrality as the basis to conduct targeted attacks. Specifically, removal of nodes with high betweenness centrality and high weighted out-degree centrality in the earlier stages is considered as efficient approaches to impair the system connectivity, thereby simulating extreme conditions. While betweenness centrality and out-degree centrality are selected as an example to implement targeted attacks in this paper, the study can be extended to other guiding metrics as well.
\begin{table}
	\centering
	\caption{Types of attack strategies}
	\label{Table:Typesofattack}
	\begin{tabular}{|c|c|c|c|}
		\hline
		\backslashbox{Strength}{Information} & \textbf{Random} & \textbf{Target} \\
		\hline
		\textbf{Complete} & complete random & complete target \\
		\hline
		\textbf{Partial} & partial random & partial target \\
		\hline
	\end{tabular}
\end{table}
\subsection{Complete attack}
The most trivial type of attack scenario would be the situation where the attacker is capable of disabling a process completely, referred to as a complete attack. This can be caused by extreme natural events or powerful malicious attackers. In WHFGT, complete attacks can be represented by removal of corresponding nodes entirely, within a single stage. Based on the knowledge level of the attacker, complete attacks can be further categorized into complete random and complete targeted attacks. Specifically, for a complete random attack, it is assumed that the attacker is not aware of the internal configuration of the IUN and hence the targeted nodes (processes) will be attacked randomly. Under this situation, nodes are selected randomly for removal at each stage of the percolation and the percolation is continued until the graph is left with two nodes. 
Obviously, if a child node (process) is dependent on its only parent node (process), disabling the parent node (process) will lead to obliteration of the child node. For instance, \enquote{ the node representing the process of  transport water from WTP to the residential community}  will be removed once the node corresponding to the process of \enquote{treat water in WTP} is out of the graph. On the other hand, for child nodes with multiple parent nodes, removing some (not all) of its parent nodes will lead to a certain degree of degradation, which is quantified by the out-degree weight reduction. Out-degree weights of a node indicate the amount of dependency the neighbors have on the node, whereas the in-degree weights of a node signify the extent of dependency the node has on its neighbor. Specifically, the decrease of out-degree weights is assumed to be in proportion to the decrease of in-degree weights. In other words, for a particular node (process), with less inputs from parent nodes (processes), its performance of supporting other nodes (processes) will be degraded accordingly. When the out-degree weight of a node decreases to a pre-defined threshold, namely critical quality of functionality (QoF), it will be removed from the graph. The practical interpretation of this rule is that when a process is unable to support certain follow-up services, it will be considered as dysfunctional. For instance, when the process of \enquote{deliver electricity to customers 1-5} is degraded to \enquote{deliver electricity to only customer 1} due to attacks, the relevant infrastructure might be considered as dysfunctional and the temporary shutdown \& maintenance actions may be needed. Note that the critical QoF value can be assigned flexibly, based on the specific processes and situations. 

\begin{algorithm}[h!]
	\caption{Complete attack}
	\begin{algorithmic}[1]
		\renewcommand{\algorithmicrequire}{\textbf{Input:}}
		\renewcommand{\algorithmicensure}{\textbf{Output:}}
		\REQUIRE Graph $G$ with $V$ nodes.
		\ENSURE  List of robustness metrics corresponding to various percolation stages   
		\\ \textit{LOOP Process}
		\WHILE{$|V|>2$}
		\STATE \textit{nodeselected}: Select a node randomly or based on the order of important scores.
		\STATE \textit{childnodes}: Find all child nodes of \textit{nodeselected}.
		\STATE remove  \textit{nodeselected} from $G$
		\FOR {\textit{child} in \textit{childnodes}}
		\STATE  \textit{indeg}: Find weighted in degree of \textit{child} node
		\STATE  \textit{outdeg}: Find weighted out degree of \textit{child} node 
		\STATE \textit{Degradationratio}$=\frac{\text{current indeg of child }}{\text{indeg of child before percolation}}$
		\STATE \textit{outdegweights}: Decrease the out edges weight of \textit{child} by \textit{Degradationratio}
		\IF {( \textit{indeg}$=0$) or (outdegweights $<$ Quality of Function)}
		\STATE remove  \textit{child} from $G$
		\ENDIF
		\ENDFOR
		\STATE compute robustness metrics
		\ENDWHILE 
		\RETURN list of robustness metrics corresponding to various percolation stages 
	\end{algorithmic} 
\end{algorithm}

Furthermore, considering the frequent occurrences of malicious cyber attacks around the globe, it is increasingly vital to address the situation wherein the attacker could obtain insights on the entire system and launch targeted attacks. Targeted attacks aim to remove nodes with higher weighted out-degree/betweenness centrality in the earlier stages of percolation. In other words, processes with higher importance are targeted in the earlier stages of percolation. The complete targeted attack is similar to its random counterpart except that the nodes are selected on the basis of predefined importance score, determined by weighted out-degree/betweenness centrality. During targeted attacks, nodes are removed in descending order of importance. The detailed procedure of complete attack is shown in Algorithm 1. The robustness metrics indicated in the output of Algorithm 1 will be elaborated in section 4. Again, it should be noted that the word \enquote{attack} used in this paper does not necessarily denote malicious cyber/physical destruction, but also represents natural disasters that may lead to serious damage.
\subsection{Partial attack}
The complete attack is designed for simulating the scenarios where the attack is sufficiently powerful to destroy a process completely, which results in loss of the entire process. To examine the robustness of the graph against less severe contingencies, we propose a new type of attack strategy referred to as partial attack. The key difference between partial and complete attack strategies is that during percolation of a partial attack, the attacked nodes will only experience out-degree depletion until the critical QoF value is violated. In other words, unlike complete attacks where the attacked process is disabled unconditionally, partial attacks will only lead to a certain level of degradation. Without loss of generality, the degradation level is set to 20\% in this paper, which can be adapted to other levels based on specific conditions. On the other hand, similar to complete attacks, partial attacks can be further subdivided into two types depending upon the knowledge level of the attacker, namely partial random attacks and partial targeted attacks. For partial random attacks, the attacked nodes are selected randomly, while nodes with higher weighted out-degree/betweenness centrality will be targeted in the early stages during partial targeted attacks. Algorithm 2 describes the complete procedure of a partial attack. It should be noted that partial attack analysis only applies to our proposed weighted hetero functional graphs, since partial degradation of weights can be incorporated. Traditional HFGT framework cannot accommodate these realistic attack scenarios.

\begin{algorithm}[h!]
	\caption{Partial attack}
	\begin{algorithmic}[1]
		\renewcommand{\algorithmicrequire}{\textbf{Input:}}
		\renewcommand{\algorithmicensure}{\textbf{Output:}}
		\REQUIRE Graph $G$ with $V$ nodes.
		\ENSURE  List of robustness metrics corresponding to various percolation stages   
		\\ \textit{LOOP Process}
		\WHILE{$|V|>2$}
		\STATE \textit{nodeselected}: Select a node randomly or based on the order of important scores.
		\STATE \textit{childnodes}: Find all child nodes of \textit{nodeselected}.
		\STATE \textit{outdegweights}: Decrease the out edges weight of \textit{nodeselected} by random quantity.
		\IF {( \textit{outdegweights} $<$ Acceptable service)}
		\STATE remove  \textit{nodeselected} from $G$
		\ENDIF
		\FOR {\textit{child} in \textit{childnodes}}
		\STATE  \textit{indeg}: Find weighted in degree of \textit{child} node
		\STATE  \textit{outdeg}: Find weighted out degree of \textit{child} node 
		\STATE \textit{Degradationratio}$=\frac{\text{current indeg of child }}{\text{indeg of child before percolation}}$
		\STATE \textit{outdegweights}: Decrease the out edges weight of \textit{child} by \textit{Degradationratio}
		\IF {( \textit{indeg}$=0$) or (outdegweights $<$ Quality of Function)}
		\STATE remove  \textit{child} from $G$
		\ENDIF
		\ENDFOR
		\STATE compute robustness metrics
		\ENDWHILE 
		\RETURN list of robustness metrics corresponding to various percolation stages
	\end{algorithmic} 
\end{algorithm}

\section{Numerical Study}

In this section, we implement weighted Hetero-functional graph theory based modeling of IUN and evaluate its robustness. To begin with, the system network configuration is described, along with its effects on the process relation matrix ($P_R$). After introducing several robustness metrics, the IUN robustness is quantified by analyzing the impacts of four types of attack scenarios. Finally, the application of proposed robustness assessment method in optimizing dependency relations among infrastructures for robustness enhancement is demonstrated.

\subsection{System Network Configuration}
\begin{figure}[!ht]
	\centering
	\includegraphics[height=12cm, width=16cm]{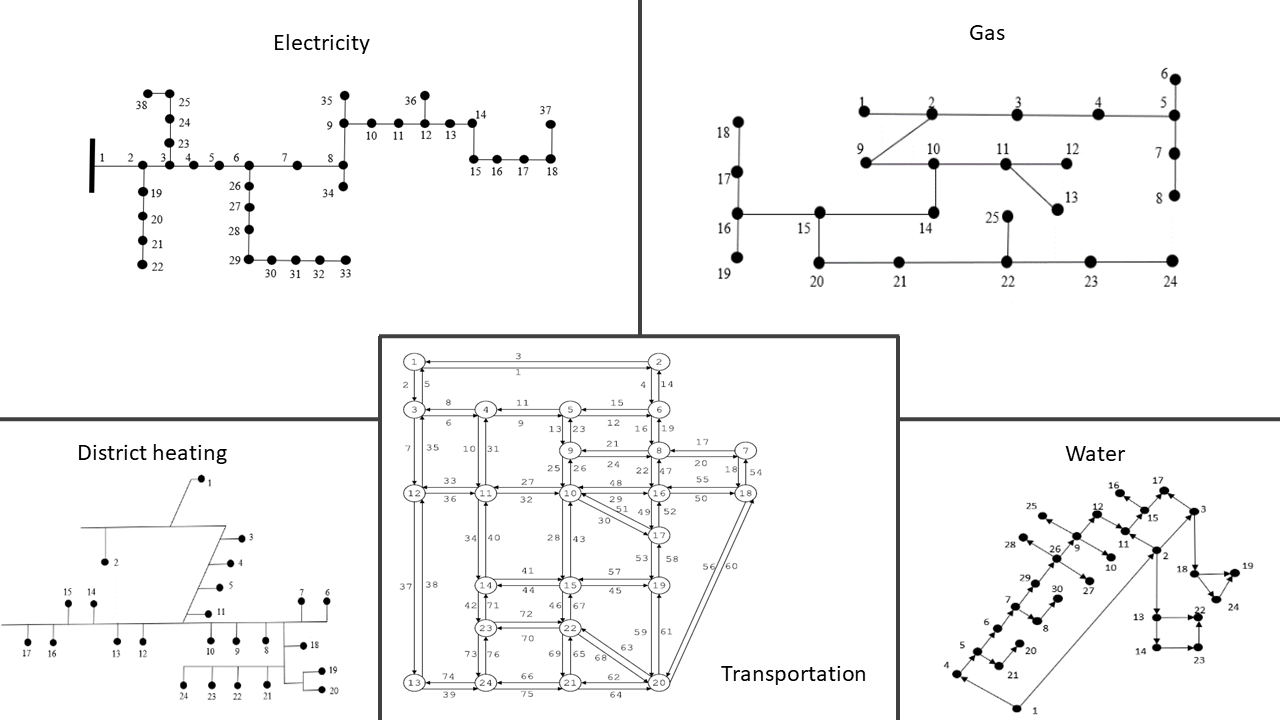}
	\caption{Standard test networks corresponding to power, water, transport, gas and heating networks}
	\label{Fig:Testnetwork}
\end{figure}

Fig. \ref{Fig:Testnetwork} shows the integrated urban network of interest. This network comprises of an electric power system \cite{Singh2007}, a gas network \cite{Sundar2019}, a water system \cite {Sherali2001}, a road transportation network \cite{Tao2018}, and a district heating network \cite{Zheng2017}. Further, Table \ref{Table:Testnetworkdesc} tabulates the locations of various infrastructure facilities within the networks, including 1 gas-fired power plant (GPP), 1 solar power plant (SPP), 2 water treatment plants (WTP), 3 heating plants (HPL), 2 gas stations (GS), 1 gas-driven CHP, 2 industry parks (IP1-IP2), 3 commercial zones (CZ1-CZ3), 6 residential zones (RZ1-RZ6), 1 water infrastructure repair facility (WIRF), 1 power infrastructure repair facility (PIRF) and 1 natural gas infrastructure repair facility (GIRF). To ensure the proper operation of these facilities, 121 processes are required in total. In particular, electricity supplies of industry parks, commercial zones and one residential zone (RZ1) depend on three electric power sources, including the gas-driven power plant, CHP and the solar plant. The complete process relation matrix is presented in \cite{ProcessMatrix1}.

For this system network configuration, the relation matrix needs to be extended to incorporate the additional dependencies among processes due to the shared path of energy/service deliveries. For instance, the successful execution of the process that transmits power from GPP (located at node 5 in electric power network) to WIRF (located at node 1 in electric power network) depends on the successful execution of process that transmits power from GPP (located at node 5 in electric power network) to WTP1 (located at node 2 in electric power network). This is because, the electricity delivery path between GPP (node 5) and WTP1 (node 2) is a part of the only path for electricity transmission from GPP (node 5) to WIRF (node 1). Similarly, the effects of other network constraints on process dependencies are integrated into the process relation matrix.  

\begin{table}[h!]
	\centering
	\captionsetup{width=.85\textwidth}
	\caption{Locations of Entities in Various Networks}
	\label{Table:Testnetworkdesc}
	\begin{tabular}{|c|c|c|c|c|c|c|}
		\hline
		\backslashbox{Utility}{Description} & Entity & Location & Entity & Location & Entity & Location   \\ 
		\hline
		& WIRF & Node 1 & CZ1 & Node 9 & RZ2 & Node 17\\
		& WTP1 & Node 3 & WTP2 & Node 11 & RZ3 & Node 19 \\
		& IP1 & Node 4 & CZ2 & Node 12 & RZ4 & Node 20 \\
		Power & GPP & Node 5 & RZ1 & Node 13 & GS2 & Node 21 \\
		& IP2 & Node 6 & GIRF & Node 14 & RZ5 & Node 22 \\
		& GS1 & Node 7 & SPP & Node 15 & CHP1 & Node 23 \\
		& PIRF & Node 8 & CZ3 & Node 16 & RZ6 & Node 24 \\
		\hline   
		& WIRF & Node 1 & CZ1 & Node 9 & RZ2 & Node 17 \\
		& HPL1 & Node 2 & HPL2 & Node 10 & HPL3 & Node 18 \\
		& WTP1 & Node 3 & WTP2 & Node 11 & RZ3 & Node 19 \\
		Water & IP1 & Node 4 & CZ2 & Node 12 & RZ4 & Node 20 \\
		& GPP & Node 5 & RZ1 & Node 13 & GS2 & Node 21 \\
		& IP2 & Node 6 & GIRF & Node 14 & RZ5 & Node 22 \\
		& GS1 & Node 7 & SPP & Node 15 & RZ6 & Node 24 \\
		& PIRF & Node 8 & CZ3 & Node 16 &  & \\
		\hline
		& GS1 & Node 1 & HPL3 & Node 8 & RZ3 & Node 14 \\
		& HPL1 & Node 2 & CHP1 & Node 10 & RZ4 & Node 15 \\
		Gas & GPP & Node 5 & RZ1 & Node 12 & RZ5 & Node 17 \\
		& HPL2 & Node 6 & RZ2 & Node 13 & RZ6 & Node 19 \\
		\hline
		& WIRF & Node 1 & CZ2 & Node 12 & RZ3 & Node 19 \\
		& IP1 & Node 4 & RZ1 & Node 13 & RZ4 & Node 20 \\
		Heat & IP2 & Node 6 & GIRF & Node 14 & RZ5 & Node 22 \\
		& PIRF & Node 8 & CZ3 & Node 16 & CHP1 & Node 23 \\
		& CZ1 & Node 9 & RZ2 & Node 17 & RZ6 & Node 24 \\
		\hline
		& WIRF & Node 1 & PIRF & Node 8 & HPL3 & Node 18 \\
		& HPL1 & Node 2 & HPL2 & Node 10 & GS2 & Node 21 \\
		Transport & WTP1 & Node 3 & WTP2 & Node 11 & CHP1 & Node 23 \\
		& GPP & Node 5 & GIRF & Node 14 &  & \\
		& GS1 & Node 7 & SPP & Node 15 &  &  \\
		\hline
	\end{tabular}
\end{table}


\subsection{Metrics for robustness evaluation}
Robustness of IUN in the present work is studied through the percolation process, where nodes of the network are removed in stages and the connectivity of the residual network is assessed at each stage, until the graph is completely disconnected. In a high level definition, the connectivity of the residual graph after node removal represents the robustness of the graph towards that node attack. In general, higher connectivity of the residual graph indicates enhanced robustness. The conventional indices for quantification of graph connectivity include Largest Connected Component (LCC) and the Number of Connected Components (NCC) \cite{Ahmed2011}. Here, a component denotes a subgraph, in which any two vertices are connected to each other by paths and is connected to no additional vertices in the graph. Specifically, LCC denotes the size of the largest component where every node is at least connected to one other node, while NCC represents the number of connected components in the entire graph. Furthermore, we use flow robustness (FR) to quantify robustness, from the components standpoint \cite{alenazi2015comprehensive}. It captures the ability of the nodes to communicate with each other in all the clusters and hence characterizes the overall reachability of the graph. Unlike LCC that only accounts for the largest connected component, FR incorporates the number of nodes in all components of the residual graph. The FR  metric corresponds to:
\begin{equation}
FR = \frac{\sum_{i}|C_{i}|(|C_{i}|-1)}{N(N-1)}    
\end{equation}
where $C_{i}$ is the number of nodes in component $i$ and $N$ denotes the total number of nodes in the original graph before attacks. As seen, FR represents the degradation at a global level by monitoring the connectivity situation of all components and hence reveals global robustness. 

However, the three indices described earlier are inherently incapable of incorporating weights of edges. Therefore, a new index, namely Service robustness (SR), that leverages the weights of edges is designed as:

\begin{equation}
SR=\frac{ \sum_{i} Q^{a}_{i}}{\sum_{i} Q_{i}} 
\end{equation}
where $Q^{a}_{i}$ denotes the weighted out-degree of node $i$ after each stage of attack and $Q_{i}$ represents the total weighted out-degree of all nodes in the original graph without attacks. The SR index not only infers connectivity of the graph, but also incorporates the weights of edges. Since the weights of edges quantify the dependencies between source and target nodes, integrating them into the robustness analysis would aid in providing a more precise evaluation of impacts of attacks. 

To summarize, it is obvious that higher values of LCC, FR and SR indicate a more robust network, as they quantify reachability and connectivity from various dimensions. However, the relationship between the NCC and robustness is not straightforward. Specifically, at the initial stages of decomposition process of the graph, the value of NCC may increase since the entire graph is decomposed into several sub-graphs (components). After a certain stage, the value of NCC decreases gradually, as components start to vanish into individual nodes. In the next sub-section, we examine the robustness of the graph described earlier by observing the trajectories of the four robustness metrics along the percolation stages with different types of attacks.  




\subsection{IUN Robustness Analysis}
The robustness of IUN is examined by evaluating the robustness metrics, i.e., $LCC$, $NCC$, $SR$ and $FR$, after each stage of percolation. The impacts of four attack strategies, i.e., complete random attack, complete targeted attack, partial random attack and partial targeted attack, are examined and compared in the following. The attacked node selections in targeted attack are carried out based on betweenness and weighted out-degree. Additionally, Monte-Carlo simulation approach is used here to evaluate the effects of random attack strategies on robustness metrics. Here, the random selections of attacked nodes are repeated 10,000 times.   

\subsubsection{Complete attack}
\begin{figure}[!ht]
	\centering
	\includegraphics[scale=0.3]{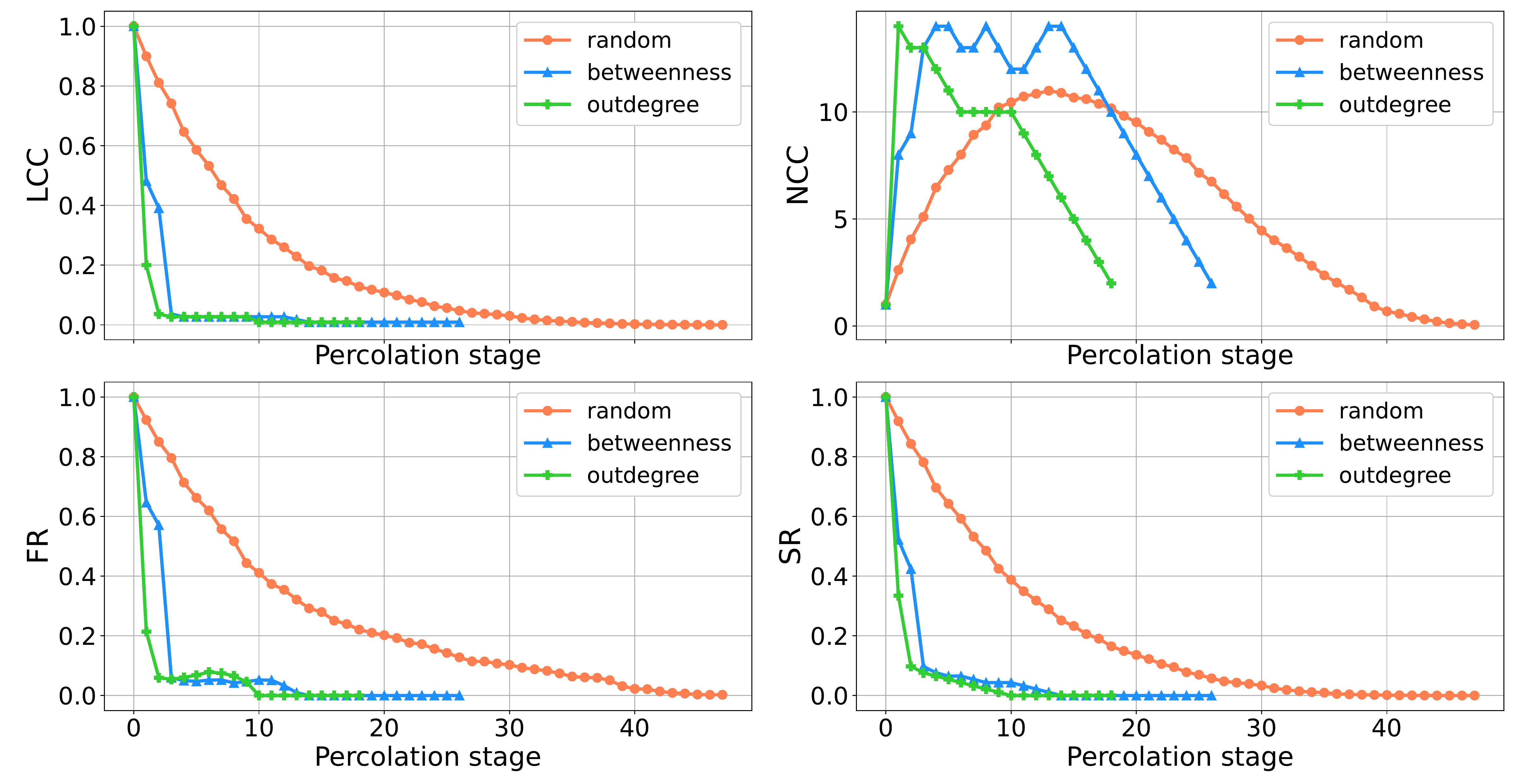}
	\caption{Trajectory of robustness metrics across various stages of percolation with complete attack}
	\label{Fig:Completeattack}
\end{figure}

As mentioned earlier, in a complete attack, the attacked nodes are removed completely once the attack is imposed. The trajectories of the four robustness metrics are depicted in Fig. \ref{Fig:Completeattack}. As expected, LCC decreases along with the progress of percolation, as node removals will decompose the graph into various components with a smaller number of nodes. On the other hand, the NCC increases in the initial stages and starts declining after a certain stage. This is because, initially, the entire graph is fragmented into several sub-graphs that lead to an increasing number of connected components. After a certain stage (stage $14$ in this case), those sub-graphs will be further decomposed into individual nodes, decreasing the value of NCC. The global metric FR also drops exponentially but with a lower decay rate compared to LCC, which implies the slower degradation of connectivity (weighted or unweighted) among various components across the entire graph (compared to the local connectivity in the largest component). The trajectory of SR shows a gradual decline as well, which captures the diminishing mutual dependencies among nodes. 

To demonstrate the effects of attackers' knowledge about the system, we further illustrate the trajectories of robustness metrics with imposed betweenness- and weighted outdegree-based targeted attacks along the percolation stages, as depicted by blue and green curves in Fig. \ref{Fig:Completeattack}, respectively. Like the results associated with complete random attacks, the general patterns of trajectories of robustness metrics correspond to the trend of gradual decline except for NCC. The NCC metric initially increases abruptly followed by fluctuations and eventually settles in the final stages. Additionally, the value of FR has a lower decay rate compared to LCC value, which indicates a certain level of connectivity among components even with few nodes in the largest component. Additionally, compared to a random attack, the degradation of the system robustness is substantially higher with targeted attacks. Specifically, it takes $12$ and $10$ stages to disconnect the graph entirely if betweenness and weighted out-degree are selected as the basis to launch targeted attacks, compared to an average of $40$ stages required to completely disconnect the graph with imposed random attacks. To further quantify the distinctions between random and targeted attacks, the numbers of sequential attacks required for degrading the robustness metrics to certain levels are tabulated in Table \ref{Table:Completeweighted}. The results illustrate that targeted attacks require fewer steps to reach the same level of degradation as random attacks. For instance, achieving $50\%$ degradation in FR only takes 3 and 1 stages for betweenness- and weighted outdegree-based targeted attacks, respectively, while 10 steps are needed to attain the same level of degradation with random attacks. In other words, this test demonstrates the value of securing system information, as attackers with system information are capable of causing severe destruction rapidly. Furthermore, comparing two types of targeted attacks, we can observe that the weighted outdegree-based targeted attacks outperform betweenness-based targeted attacks for all robustness metrics. This implies that weighted outdegree centrality could provide a more precise indication of nodes importance in this test.    



\begin{table}[h!]
	\centering
	\captionsetup{width=.85\textwidth}
	\caption{ Number of attacks for various degradation level corresponding to random and targeted attacks with complete attack strategy RN: Random attack BW: Betweenness based targeted attack OD: Out-degree based targeted attack}
	\label{Table:Completeweighted}
	\begin{tabular}{|c|c|c|c|c|c|c|c|c|c|}
		\hline
		Degradation Level & \multicolumn{3}{c|}{20\% } & \multicolumn{3}{c|}{50\%} & \multicolumn{3}{c|}{80\%}   \\ 
		\cline{1-10}
		\backslashbox{Metric}{Attack}& RN & BW & OD & RN & BW & OD & RN & BW & OD \\
		\hline
		LCC  & 3 & 1 & 1 & 7 & 1 & 1 & 15 & 3 & 2  \\
		\hline
		FR & 4 & 1 & 1 & 10 & 3 & 1 & 22 & 3 & 2  \\
		\hline
		SR & 3 & 1 & 1 & 8 & 2 & 1 & 17 & 3 & 2  \\
		\hline
	\end{tabular}
\end{table}

\subsubsection{Partial attack}
\begin{figure}[!ht]
	\centering
	\includegraphics[scale=0.3]{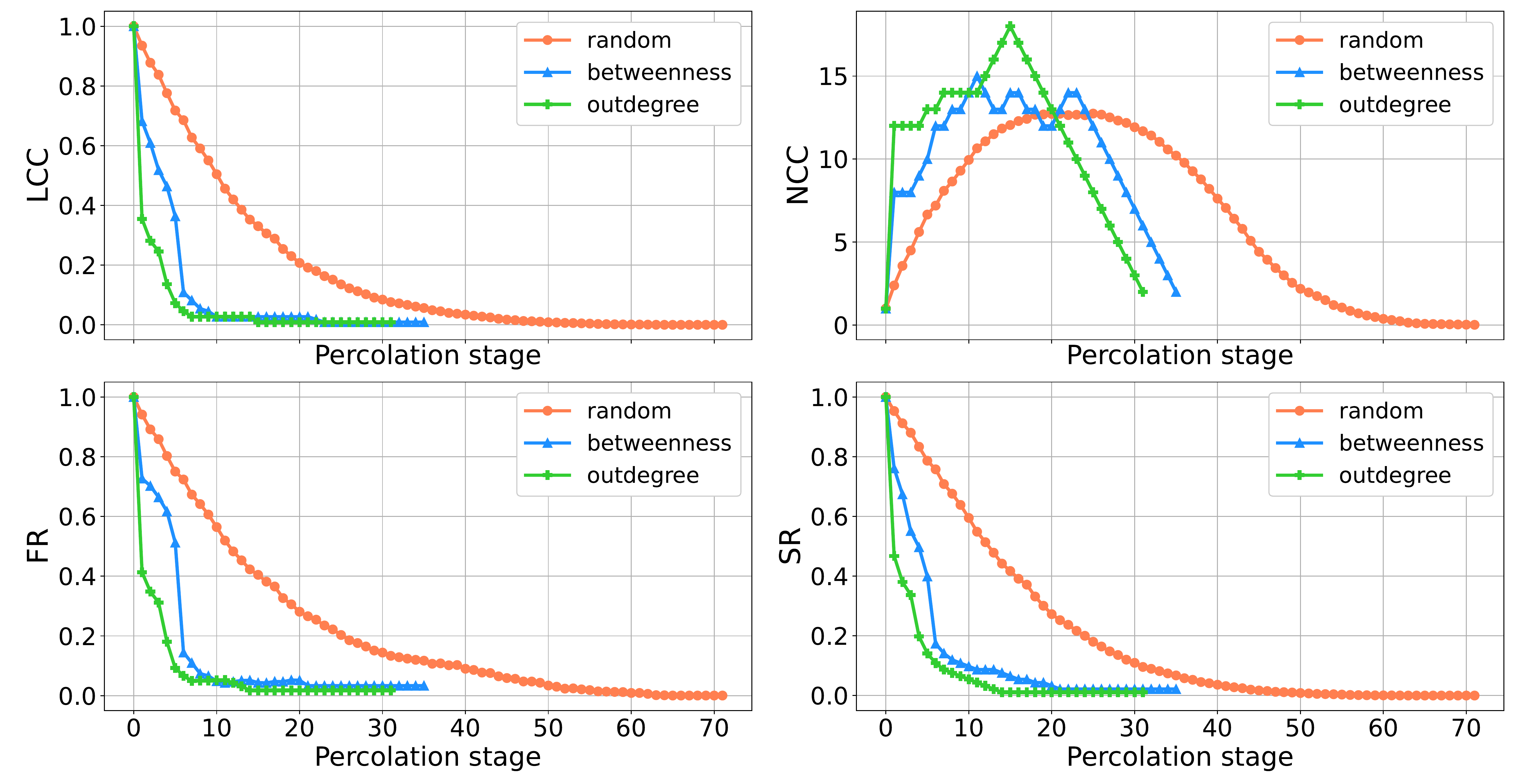}
	\caption{Trajectory of robustness metrics across various stages of percolation with partial attack.}
	\label{Fig:Partialattack}
\end{figure}

In partial random attacks, attacked nodes are selected randomly with a certain level of degradation at each stage of percolation. The orange curves in Fig. \ref{Fig:Partialattack} depict the trajectories of robustness metrics. The general trends in Fig. \ref{Fig:Partialattack} are similar to the results of a complete random attack. Specifically, the metrics of LCC, FR and SR decrease abruptly at the beginning followed by a moderate decline, while the index of NCC increases initially and decreases after a certain stage. These observations are consistent with the results from complete attack case described earlier. 

\begin{figure}[!ht]
	\centering
	\includegraphics[height=6cm, width=13cm]{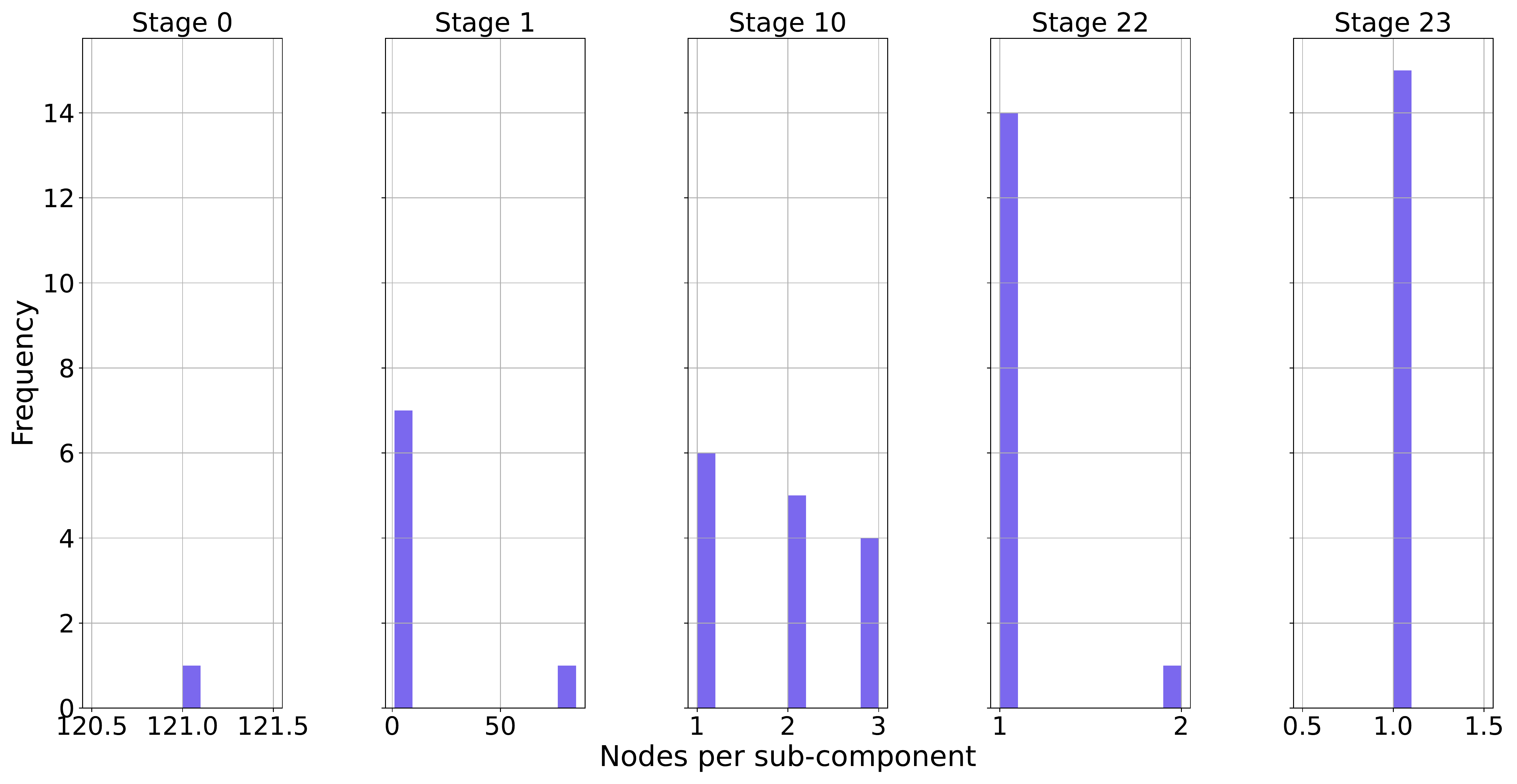}
	\caption{Histogram of nodes per sub-component/cluster across different stages of percolation in partial targeted attack}
	\label{Fig:connectedcomp}
\end{figure}

We also illustrate the trajectories of robustness metrics along the percolation stages with partial targeted attacks, as depicted in blue and green curves in Fig. \ref{Fig:Partialattack}. As seen in the figure, targeted attacks lead to a more abrupt decrease in LCC, FR and SR. To further illustrate the graph fragmentation process, Fig. \ref{Fig:connectedcomp} depicts the histograms of number of nodes in each component at representative stages of partial targeted attack. The rapid decrease of node number in each component at the beginning can be explicitly observed, while the component decomposition process significantly slows down after stage $10$. The results correspond to the conclusion drawn from Fig. \ref{Fig:Partialattack}, which again signifies the remarkable ability of targeted attacks in decimating the network rapidly. Additionally, the notable gap between partial random and targeted attack strategies can be observed in Table \ref{Table:Partialattack}, which tabulates the required numbers of attack stages to reach certain degradation levels. As expected, it takes fewer steps for targeted attacks to fragment the graph into a certain level. In a nutshell, the general trend of robustness metrics are similar in both complete and partial attacks. The distinction between complete and partial attacks is compared with the distinction between targeted and random attacks in the following subsection.

\begin{table}[h!]
	\centering
	\captionsetup{width=.85\textwidth}
	\caption{Number of attacks for various degradation level corresponding to random and targeted attacks with partial attack strategy. RN: Random attack BW: Betweenness based targeted attack OD: Out-degree based targeted attack}
	\label{Table:Partialattack}
	\begin{tabular}{|c|c|c|c|c|c|c|c|c|c|}
		\hline
		Degradation Level & \multicolumn{3}{c|}{20\% } & \multicolumn{3}{c|}{50\% } & \multicolumn{3}{c|}{80\%}   \\ 
		\cline{1-10}
		\backslashbox{Metric}{Attack} & RN & BW & OD & RN & BW & OD & RN & BW & OD \\
		\hline
		LCC  & 4 & 1 & 1 & 11 & 4 & 1 & 23 & 6 & 4  \\
		\hline
		FR & 4 & 1 & 1 & 13 & 6 & 1 & 30 & 6 & 4  \\
		\hline
		SR & 5 & 1 & 1 & 13 & 4 & 1 & 26 & 6 & 4  \\
		\hline
	\end{tabular}
\end{table}
\subsubsection{Complete random attack versus partial targeted attack}

Evidently, complete, targeted attacks cause more severe damage than partial, random attacks, as demonstrated in Tables \ref{Table:Completeweighted} and \ref{Table:Partialattack}. In this section, the comparison between the impacts of a complete random attack and a partial targeted attack is conducted to evaluate the merits of securing system information and infrastructure hardening.

\begin{figure}[!ht]
	\centering
	\includegraphics[height=11cm, width=15.5cm]{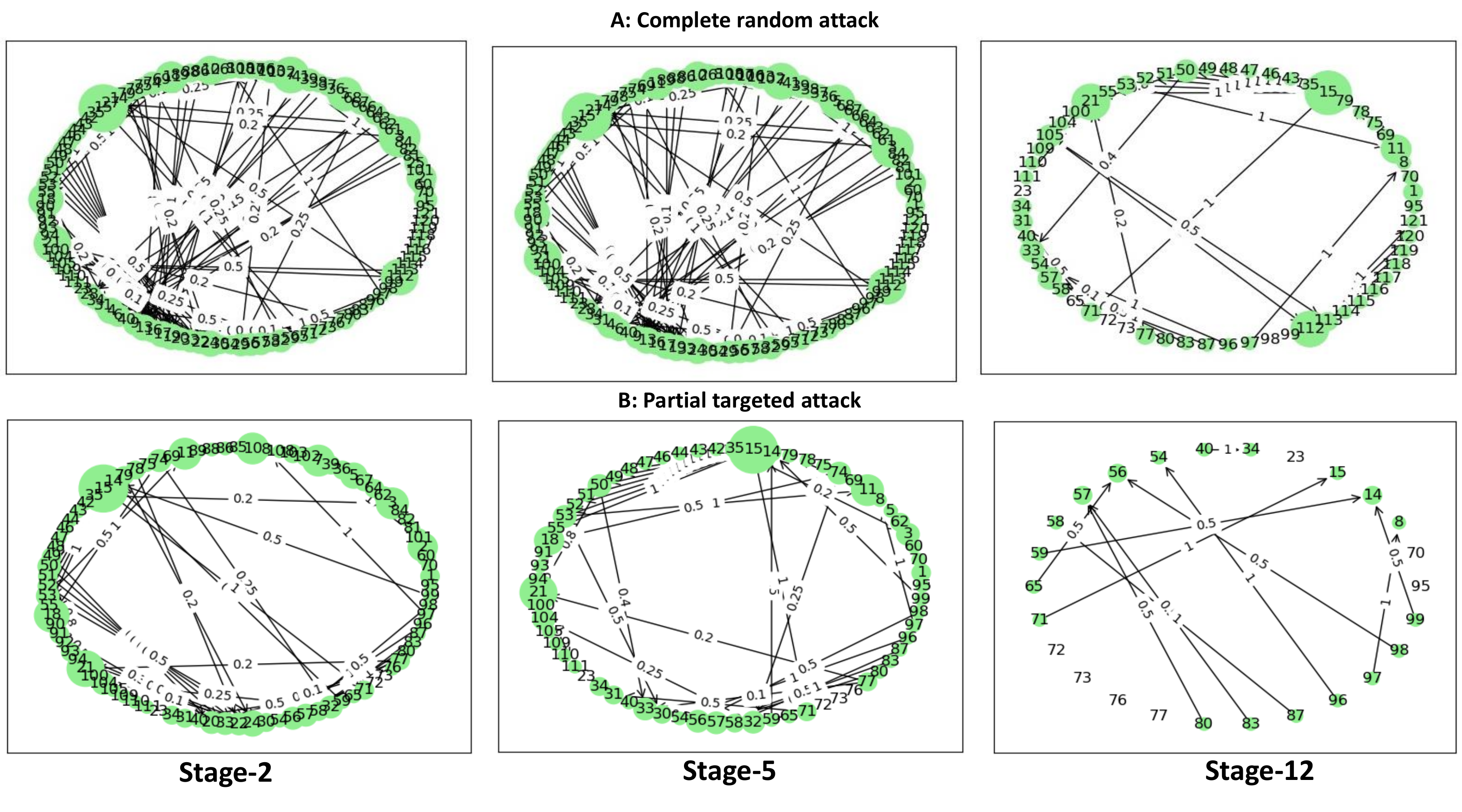}
	\caption{Snapshots of graph across representative stages of percolation with complete random attack and partial targeted attack.}
	\label{Fig:stagescomppartial}
\end{figure}

To begin with, from Tables \ref{Table:Completeweighted} and \ref{Table:Partialattack}, we can observe that the distinction between random and targeted attacks is more significant, compared to the distinction between complete and partial attacks. For instance, the gaps between a partial random attack and a complete random attack, along with a partial targeted attack and a complete targeted attack in term of the required stages to cause $80\%$ degradation in $FR$ are $8$ and $2$ stages, respectively. Whereas, the gaps between a partial random attack and a partial targeted attack, along with a complete random attack and a complete targeted attack for the same goal are $26$ and $20$ stages, respectively. To further illustrate and demonstrate which factor dominates severity of attacks, strength or intelligence, Fig. \ref{Fig:stagescomppartial} depicts snapshots of the residual graph at three representative stages with a complete random attack and a partial targeted attack. It can be observed that the partial targeted attack outperforms the complete random attack in rate of node eliminations. This study reinforces the merits of securing system information for robustness enhancement.


\begin{figure}[!ht]
	\centering
	\includegraphics[height=12cm, width=8cm]{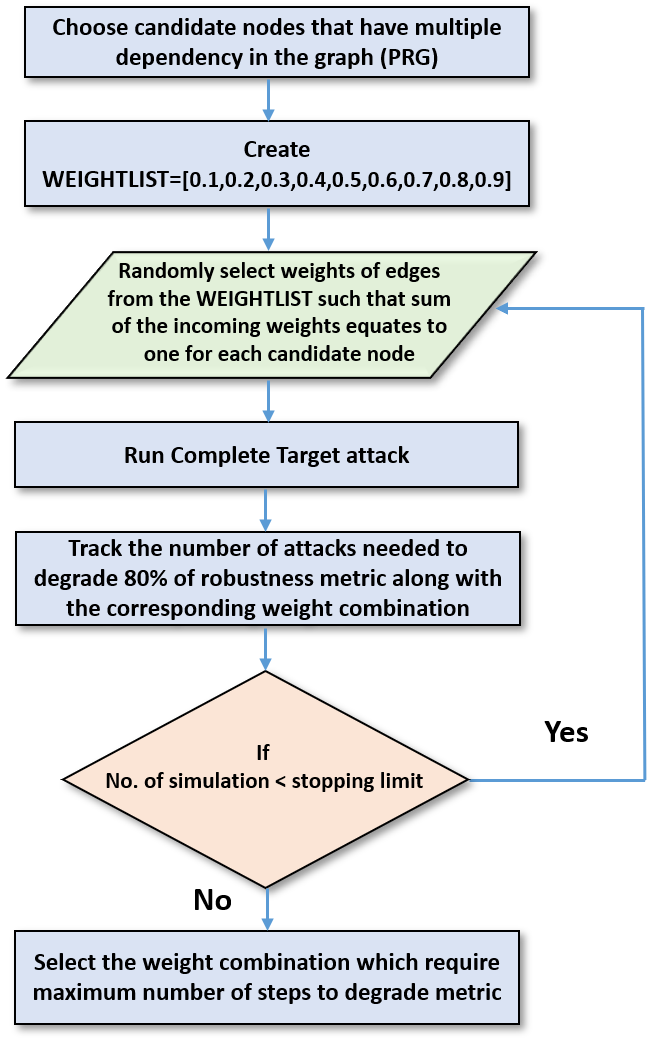}
	\caption{Flowchart depicting the enumeration }
	\label{Fig:weightalloc}
\end{figure}

\subsection{WHFGT based Optimal Dependency allocation }
Apart from identifying influential processes and evaluating robustness, the proposed framework can aid in deriving the optimal dependency between processes across infrastructures that would result in enhanced robustness. The optimal dependency (weight) allocation strategy resulting in the optimized robustness metrics can be obtained by solving an optimization problem. For example, the objective could be to maximize the number of stages needed to degrade a selected robustness index by $80\%$ (as a higher number of stages implies higher robustness). Particularly, the nodes (processes) that have multiple inputs, i.e., processes that depend on various other nodes, are the candidate nodes for this optimization problem with the weights of their incoming edges as the variables that need to be optimized. Further, the sum of the incoming edge weights for any node must be equal to one. For example, We assume that weights can only be selected from the set of \{0.0, 0.1, 0.2, 0.3, 0.4, 0.5, 0.6, 0.7, 0.8, 0.9, 1.0\}. Enumeration-based method is applied to obtain the optimal solution of this NP-hard problem with the objective of maximizing the required steps to degrade $SR$ to a certain level, considering complete targeted attacks as an example. This approach can be extended to other robustness metrics and attack strategies as well. The detailed procedure is illustrated in Fig.  \ref{Fig:weightalloc}, where weights of edges are selected randomly from a finite set, while ensuring that the sum of the incoming weights for each candidate node (nodes with multiple dependency) equates to one. Then, complete targeted attack is simulated to track the number of attacks desired for $80\%$ degradation of the selected robustness index and this process is repeated for multiple times with different edge weight combination. Finally, the weight combination which embodies the highest robustness is the desired solution. In this case, for node $3$ which represents the process ``Water treatment facility WTP1 is working properly'', the obtained optimal dependency weights are $0.3$ and $0.7$ for node $35$ (``Transit power from GPP to WTP1'') and node $36$ (``Transit power from SPP to WTP1''), respectively. 
Compared to the approach based on random assignments of weights, we can observe a $12\%$ reduction of stages required to degrade $SR$ by $80\%$. Developing more efficient analytical formulations along with effective algorithms to solve them will form a path of our future direction.

\section{Conclusion}
This paper presents, for the first time, a weighted Hetero functional graph theory based framework for modeling and resilience assessment of urban integrated infrastructure networks, including electricity, water, district heating, natural gas, road transportation networks and relevant services. Specifically, system processes are denoted by nodes and edges represent the dependencies among various processes. We conduct various types of attack simulations on the proposed weighted Hetero functional graphs and provide in-depth information about the trajectories of graph fragments. In particular, to simulate real-word scenarios, four attack scenarios, namely complete random attack, complete targeted attack, partial random attack and partial targeted attack, are implemented on the network and the resulting vulnerability is analyzed via plotting the trajectories of several robustness metrics. The outcomes of robustness analysis are used to determine the optimal dependencies for robustness enhancement. Key conclusions include: (1) complete attacks cause more damage compared to partial attacks, highlighting the need for infrastructure hardening; (2) exposing network information to attackers is likely to incur substantially more serious consequences, as targeted attacks are able to decimate graph connectivity rapidly; (3) partial targeted attack causes higher damage than the complete random attack, which indicates that securing system information is more crucial than physical hardening; (4) proper dependency (weight) allocation strategy across processes would further aid in enhancing the system robustness. 

\bibliographystyle{elsarticle-num}
\bibliography{references}

\end{document}